\documentclass[3p,onecolumn,authoryear,12pt]{elsarticle}

\usepackage{epsfig}
\usepackage{graphicx}
\usepackage{float}
\usepackage{lipsum}
\usepackage{tabularx}
\makeatletter
\def\ps@pprintTitle{%
 \let\@oddhead\@empty
 \let\@evenhead\@empty
 \def\@oddfoot{}%
 \let\@evenfoot\@oddfoot}
\makeatother
\usepackage{amssymb}
\usepackage{amsmath}
\usepackage{algorithmic}
\usepackage{array}
\usepackage{natbib}
\usepackage{hyperref}
\usepackage{parskip}

\begin{document}
%EX FIG
%\begin{figure}[ht!]
% \includegraphics[scale=0.5]{diffc.JPG}
% \caption{Subtraction algorithm performed in 20,000 second intervals}
% \label{fig:subtraction}
%\end{figure}
%%%%%%%%%%%%%%%%%%%%%%%%%%%%%%%%%%%%%%%%%%%%%%%%%%%%%%%%%%%%%%%%%%%%%%%%%%%%%
%% Frontmatter
\begin{frontmatter}


% Use the tnoteref command within \title and fnref within \author or \address for footnotes;
% use the corref command within \author for corresponding author footnotes;
% \author{Name\corref{cor1}\fnref{label2}}
% \address{Address\fnref{label3}}

\title{Model Optimization for Deep Space Exploration via Simulators and Deep Learning}

% Use optional labels to link authors explicitly to addresses:

\author{James Bird\corref{cor1}%
 \fnref{fn1}}
\ead{james\_bird@ucsb.edu}

% Url can be given like this:
% \ead[url]{http://www.elsevier.com/wps/find/authorsview.authors/latex}

\author{Kellan Colburn\corref{cor1}%
\fnref{fn2}}
\ead{kellancolburn@ucsb.edu}

\author{Linda Petzold\fnref{fn1}}
\ead{petzold@ucsb.edu}

\author{Philip Lubin\fnref{fn2}}
\ead{lubin@ucsb.edu}

\cortext[cor1]{Corresponding author}
\fntext[fn1]{Department of Computer Science, University of California Santa Barbara, 2104 Harold Frank Hall, Santa Barbara, CA 93106-9530}
\fntext[fn2]{Department of Physics, Broida Hall, University of California Santa Barbara, Santa Barbara, CA 93106-9530}

\begin{abstract}

Machine learning, and eventually true artificial intelligence techniques, are extremely important advancements in astrophysics and astronomy. We explore the application of deep learning using neural networks in order to automate the detection of astronomical bodies for future exploration missions, such as missions to search for signatures or suitability of life. The ability to acquire images, analyze them, and send back those that are important, as determined by the deep learning algorithm, is critical in bandwidth-limited applications. Our previous foundational work solidified the concept of using simulator images and deep learning in order to detect planets. Optimization of this process is of vital importance, as even a small loss in accuracy might be the difference between capturing and completely missing a possibly-habitable nearby planet. Through computer vision, deep learning, and simulators we introduce methods that optimize the detection of exoplanets. We show that maximum achieved accuracy can hit above 98\% for multiple model architectures, even with a relatively small training set. 

\end{abstract}


\end{frontmatter}

\setlength\parskip{0.5cm}
\setlength\parindent{0.5cm}

%%%%%%%%%%%%%%%%%%%%%%%%%%%%%%%%%%%%%%%%%%%%%%%%%%%%%%%%%%%%%%%%%%%%%%%%%%%%%

\section{Introduction}
This paper will address some of the challenges and possibilities of exoplanet detection and classification for future exosolar system missions. Future missions may allow for travel far outside of our solar system, as well as deep into our own solar system, where return bandwidth will be severely limited; thus, choices of which data (images in particular) are important to "send back" (\cite{Lubin2016}, \cite{Lubin2020}, \cite{Sheerin2020}). The basis for exoplanetary detection via fast interstellar travel is a combination of The Starlight Program \citep{KulKarniRelativistic2017} and recent results that show how exoplanets can be detected, and distinguished from other objects, via AI-based modeling that utilizes simulated data \citep{Bird2020}. The groundwork has been laid for an AI-based small spacecraft that can travel long distances in a short amount of time, gather information on its surroundings with minimal energy requirements, and detect exoplanets and other targets of interest with excellent accuracy. The same core technology we discuss here can be applied to a wide range of astrophysics and cosmology where subtle and often transient phenomenon are critical to retrieve in low SNR situations. In future papers we will discuss using our techniques in these other application spaces.

The major points that we will discuss and examine here are related to the accuracy of exoplanetary detection. In our foundational paper \citep{Bird2020}, we used a robust model and detection score for proof of concept. Going forward, this paper will compare a wide array of models using accuracy as our main metric to determine model strength and reliability.

\section{Previous Work}
The basis for much of our work lies in deep learning via TensorFlow \citep{Abadi2016}, as well as the expected additions, such as cuDNN \citep{ChetlurcuDNN2014} and CUDA, which allows for faster deep neural network processing via a graphics processing unit (GPU). Although the idea of direct exoplanetary detection and imaging via interstellar travel is new, astronomy has been attempting the general feat via light curves for years, and even more recently with deep learning (\cite{ShallueIdentifying2017}, \cite{ZuckerShallow2018}, \cite{CarrascoDeepLearning2018}).

For direct imaging purposes, we test a variety of robust models, including variants of each model, and analyze factors such as accuracy and computational complexity. Since deep space is uncharted territory, an extremely large training data set is not possible. Therefore, we include some simpler models to offset the possibility of having models that are too advanced for the data. The overall goal of these models is to be able to identify when a planet is present in an image, while also being capable of not mistaking other astronomical objects for planets.

For the simpler models, we will compare MobileNet \citep{MobileNet}, MobileNet V2 \citep{MobileNetv2}, DenseNet 121, 169, and 201 \citep{DenseNet}, and NASNet-Mobile \citep{NASNet}. These provide solid baseline accuracy and low computational complexity, which may prove to be beneficial for our specific needs. 
For the intricate models, we will compare NASNet-Large \citep{NASNet}, Xception \citep{Xception}, VGG 16 and VGG19 \citep{VGG}, Inception V3 \citep{Inception}, Inception-ResNet V2 \citep{InceptionResNetv2}, and ResNet 50, 50 v2, 101, 101 v2, 152, and 152 v2 \citep{ResNet}. In contrast to the simpler models listed above, the training time and complexity will increase with these. However, that process is done beforehand while the wafer satellite (wafersat) is still on Earth, so these concerns are negligible when compared to the possible gains in accuracy from the more robust models. 

These models have been tested against each other in the past to some degree. ResNet has been shown to out-perform VGG (\cite{ResNet},\cite{Canziani2016}) and even advanced Inception models \citep{ResNetBeatsVGGandInception}, while other results show all of these models being out-performed by the DenseNet and InceptionResNet architectures \citep{DenseIRNwins}. 

The structure of these models and their performance is dependent on the data that is being processed. In this case, we are training on simulated images of planets and testing on real images of planets. This concept was shown to be viable in \cite{Bird2020}; however, optimizing this process will require an in-depth look at advanced deep learning techniques and models.

\section{The Process}

\subsection{Deep Neural Network Architecture}

Deep neural networks, including those used for object detection, begin by deconstructing images into pixel-based groupings that constitute an input layer. This layer, along with the hidden layer(s) and output layer, is comprised of smaller entities called neurons. Each layer of neurons is connected to the next via weights, which are learned through a training process. Gradient descent is a powerful and widely used method that allows us to minimize the cost function in order to get the most effective learning process. By taking the negative gradient, we minimize the cost function. After all is done, we are left with a network that can take an input image and output something of interest based on the training and model parameters. A more illuminating analogy would be to treat the initial inputs (pixels, or in the case of a convolutional neural network, groupings of pixels) as an input tensor. This input tensor is then essentially acted upon by a function (the neural network), which outputs a tensor corresponding to the the categorization of the input (in our case it is a binary categorization). This function initializes with random values, and is then optimized via the methods described above such that the output tensor has the highest accuracy when identifying inputted data.

\subsection{The Setup}
As discussed in detail in \cite{Bird2020}, the simulator (SpaceEngine.org) provides us with easy access to 4K, 3-D rendered images of exoplanets. Although they are randomly generated, one could create a specific planet, or filter planets by a set of conditions in order to achieve a subset of planets that have certain traits. All models were pre-trained on ImageNet \citep{ImageNet} and fine-tuned on simulated images of exoplanets. This allowed for a robust learning experience for features, and a more specific learning experience for our data set. All models were evaluated using an AMD Ryzen Threadripper 3970X 32-Core Processor @ 3.70 GHz, 128 GB of RAM and an NVIDIA Titan RTX graphics card.

In most deep learning applications for image analysis, both training sets and testing sets contain images of the same object. In our deep learning application, the training set is taken from a universe simulator (SpaceEngine.org), and the testing images are real images of planets. Without a simulator, we would not have enough images of planets, and those planets would not constitute a large enough sub-sample of possible exoplanets. By using a physics-based simulator, we can produce an abundance of realistic novel exoplanets to train on. Then, we use real planets to test the model's accuracy. This translates directly to the wafersats process during an actual interstellar journey. Image counts for all three sets is shown below in Table 1.

\begin{table}[ht!]
\centering
 \begin{tabular}{||c c c||} 
 \hline
 Training & Validation & Testing \\ [0.5ex] 
 \hline\hline
 915 & 200 & 284 \\ [1ex] 
 \hline
 \end{tabular}
 \caption{Image count for the training, validation, and testing sets.}
\end{table}

The process being performed here is unique for two major reasons. First, the entire training set is simulated images, while the entire testing set is real images. This presents a particular challenge for neural networks, as they learn in a template space and are then tested in a real space. Second, deep space provides an enormously large variety of objects. For example, gas giants vary wildly in many ways, such as size, feature differences, surface gas formations, colors, temperature, and more. In our solar system alone, we witness two quite unique gas giants: Saturn with its rings and famous hexigon, and Jupiter with its eye and dolphin formations. Training on extremely unique objects can cause neural networks to lose their generality and under-perform. The images below in Figure \ref{fig:Image_Example} compare real images taken by NASA against simulated images.

\begin{figure}[H]
\centering
 \includegraphics[width=0.5\linewidth]{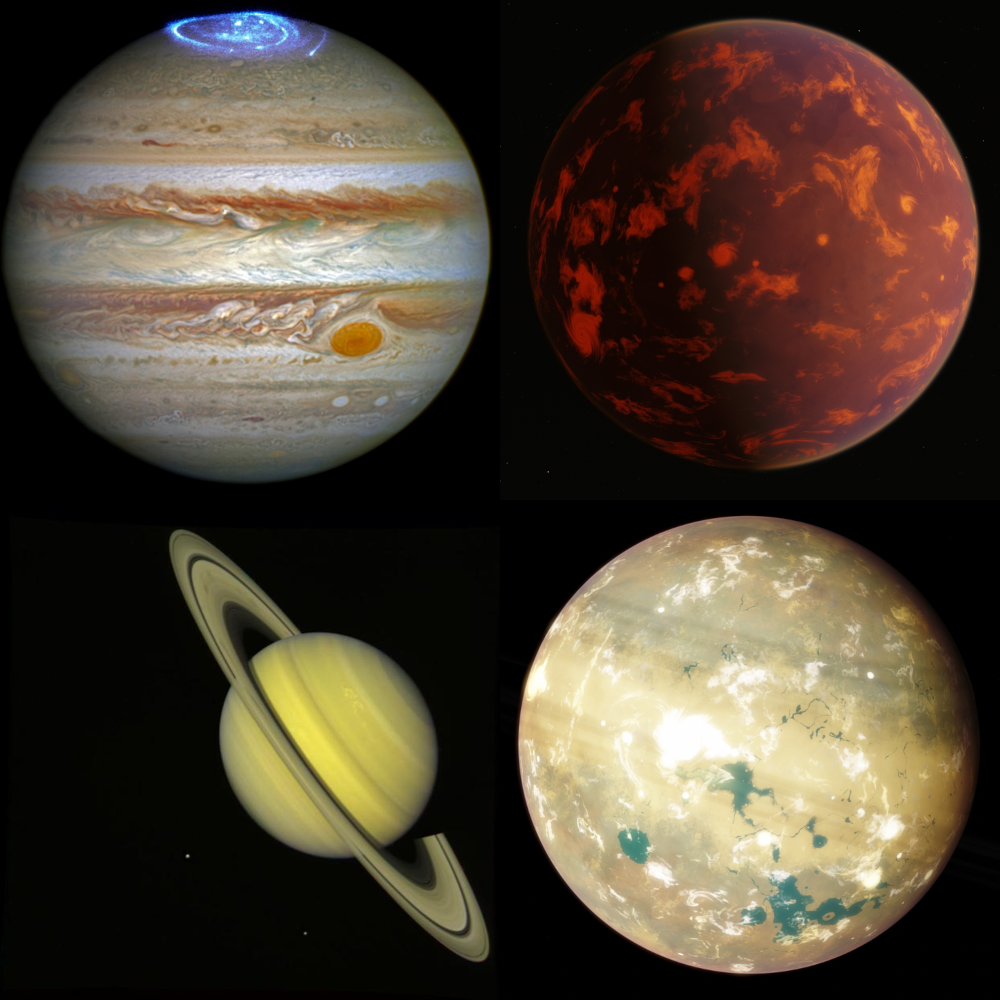}
 \caption{Two examples of real testing images are Jupiter and Saturn, seen on the left, while the right side shows examples of simulated training images.}
 \label{fig:Image_Example}
\end{figure}

%Talk about hardware used to train

\section{The Results}

For each model previously mentioned, we trained, validated, and tested the neural network in batches of five epochs each. An epoch is when the entire data set is run through the neural network once. 

\begin{table}[ht!]
\centering
 \begin{tabular}{||c c c||} 
 \hline
Model & Maximum Accuracy Achieved & Respective Epoch \\ [0.5ex] 
 \hline\hline
 VGG19 & 0.9964788556 & 5 \\ 
 \hline
 VGG16 & 0.9929577708 & 5 \& 25 \\
 \hline
 ResNet50 & 0.9894366264 & 85 \& 120 \\
 \hline
 ResNet101 & 0.9894366264 & 115 \\
 \hline
 ResNet152 & 0.9894366264 & 10 ** \\
 \hline
 MobileNet & 0.985915482 & 5 \\
 \hline
 MobileNet v2 & 0.9788732529 & 5** \\
 \hline
 DenseNet121 & 0.9788732529 & 5 \\
 \hline
 DenseNet169 & 0.9788732529 & 5 - 15 \\
 \hline
 ResNet152 v2 & 0.9753521085 & 25 * \\
 \hline
 DenseNet201 & 0.9753521085 & 5-15 \\
 \hline
 Inception v3 & 0.9683098793 & 5 \& 15 \\
 \hline
 ResNet101 v2 & 0.9647887349 & 5 \\
 \hline
 ResNet50 v2 & 0.9612675905 & 10 - 20 \\
 \hline
 NasNet-Mobile & 0.9542253613 & 10 \\
 \hline
 Inception-ResNet v2 & 0.950704217 & 5 \& 10 \\
 \hline
 Xception & 0.950704217 & 10 \\
 \hline
 NasNet-Large & 0.9154929519 & 5 \\
 \hline
\end{tabular}
\caption{Maximum epoch-based accuracy achieved for each model, ordered from highest to lowest. \textit{* denotes continued accuracy for all remaining epoch counts. ** denotes that maximum accuracy was sporadically achieved again after first occurrence.}}
\end{table}

From Table 2 above, as well as Figures \ref{fig:Results_Chart_Ordered} and \ref{fig:Max_Accuracy_Chart} below, it can be seen that VGG19 reached the highest maximum accuracy at five epochs, while VGG16 reached the second-highest maximum accuracy at both five and 25 epochs. This result is extremely interesting, as VGG variants are typically under-performing models when compared to Inception or ResNet variants. MobileNet performed extremely well, not only in maximum accuracy achieved, but also in terms of consistency. The remaining models were simply out-performed and provided no concrete reason why they should be chosen as a viable model for this specific task.

\begin{figure}[H]
\centering
 \includegraphics[width=0.8\linewidth]{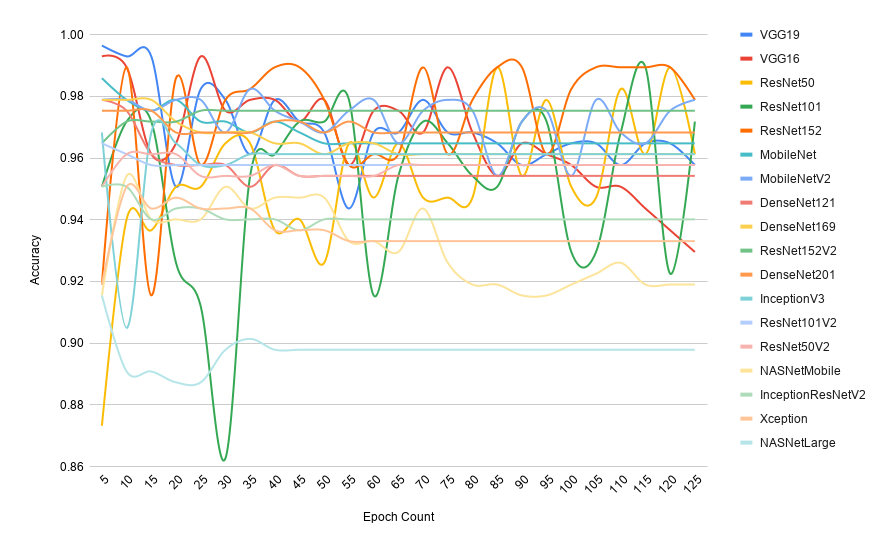}
 \caption{Accuracy of all models based on epoch count.}
 \label{fig:Results_Chart_Ordered}
\end{figure}

\begin{figure}[H]
\centering
 \includegraphics[width=0.8\linewidth]{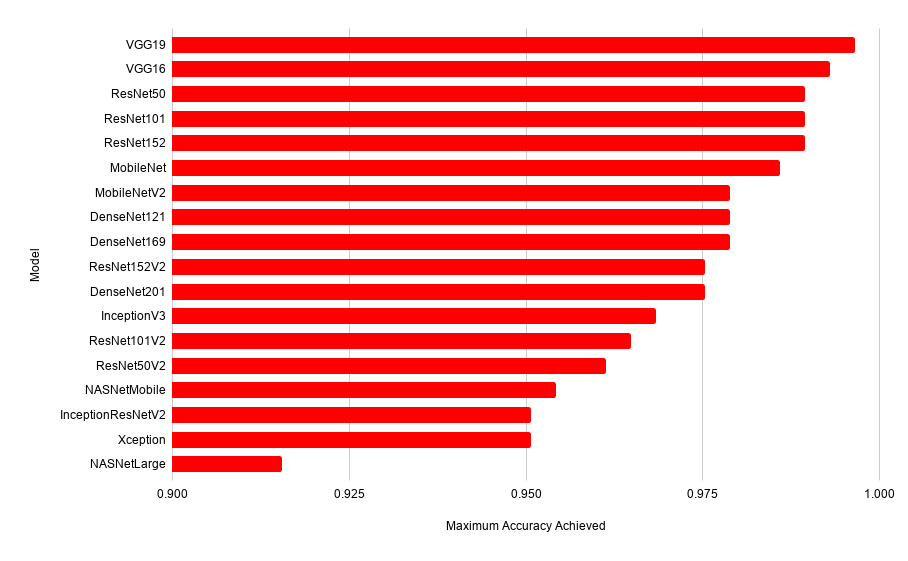}
 \caption{Accuracy of all models based on epoch count.}
 \label{fig:Max_Accuracy_Chart}
\end{figure}

Based on Figure \ref{fig:Results_Chart_98_or_above_Ordered} below, ResNet50 and ResNet101 both dipped below 88\% accuracy, and often times bounced from low to high accuracy, showing clear signs of inconsistency. Despite overall good performance, ResNet152 has large dips in the 5-20 epoch count, the main area where most models performed at their best. For these reasons, the ResNet variants were ultimately rejected as reasonable choices, as their epoch-based accuracy fluctuated too wildly.

\begin{figure}[H]
\centering
 \includegraphics[width=0.8\linewidth]{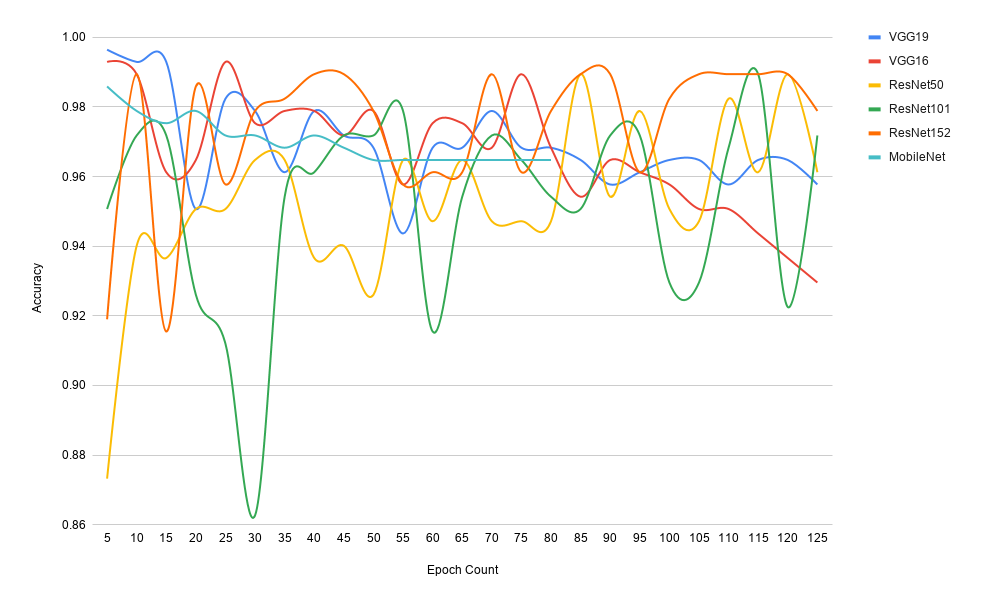}
 \caption{Accuracy of models that reached at least 98\% maximum accuracy based on epoch count.}
 \label{fig:Results_Chart_98_or_above_Ordered}
\end{figure}

One very interesting result was concerning the remaining 98\% or better maximum accuracy models, namely VGG19, VGG16, and MobileNet. Pertaining to Figure \ref{fig:Results_Chart_98_or_above_dependable} below, we can see that VGG19 and VGG16 fluctuate to some degree, while MobileNet acts as a hedge. While the dependability of VGG19 and VGG16 in certain epoch ranges is vastly superior, MobileNet grants you a consistently strong choice across all epoch ranges.

\begin{figure}[H]
\centering
 \includegraphics[width=0.8\linewidth]{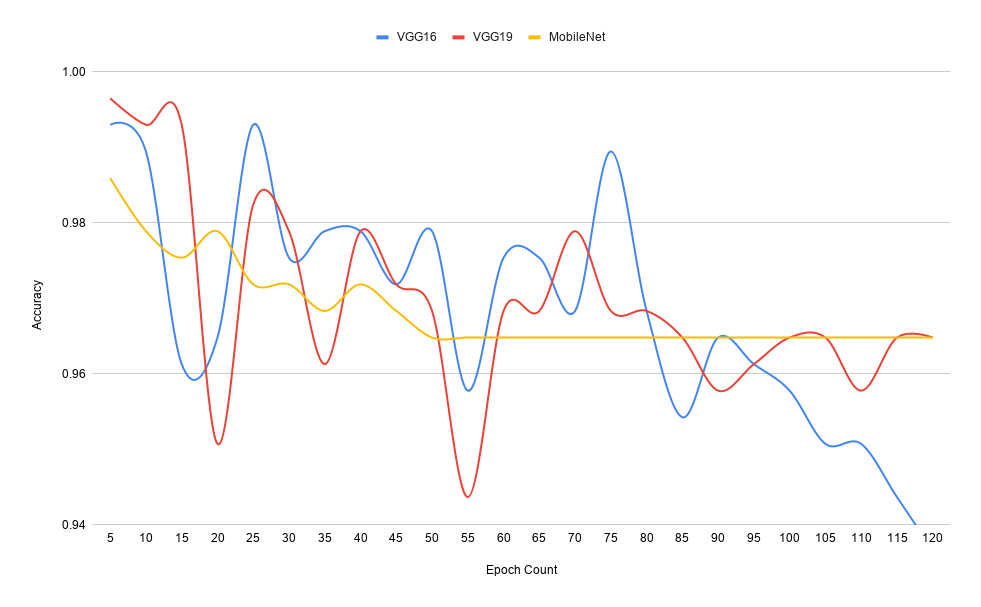}
 \caption{Accuracy of dependable models that reached at least 98\% maximum accuracy based on epoch count. These do not include the ResNet variants.}
 \label{fig:Results_Chart_98_or_above_dependable}
\end{figure}

We note that once ResNet variants are removed for their instability, every model reaches its peak accuracy at the 5-25 epoch range. Conditional on this range, the strongest models remain VGG19 and VGG16, but we can clearly see that their strength can fluctuate with small changes to epoch. Yet, MobileNet, MobileNet v2, DenseNet169, and DenseNet201 respectively perform the most consistently, while preserving most of the accuracy that we see in VGG19 and VGG16.

Some insight can be obtained by breaking down the accuracy into false negatives, which occur when a planet is present but the model does not identify it, and false positives, which occur when a planet is not present but the model identifies one. For our objective, false negatives are a much more severe error, as finding a planet and missing it is the worst possible situation. Alternatively, false positives would simply send back a picture of empty space to Earth, which would result in something mildly interesting, but nothing lost. 
Linking this information to our previous findings, ResNet variants continue to show instability, with some models having as many as 13 false negatives. Both VGG variants had no false negatives, meaning that they exhibit both extreme accuracy and reliability in detecting planets when they are actually present in the image. Lastly, we noted that DenseNet variants were very reliable in the 5-25 epoch range. In terms of false negatives, all DenseNet variants continue this stability with no false negatives.

\begin{figure}[H]
\centering
 \includegraphics[width=0.8\linewidth]{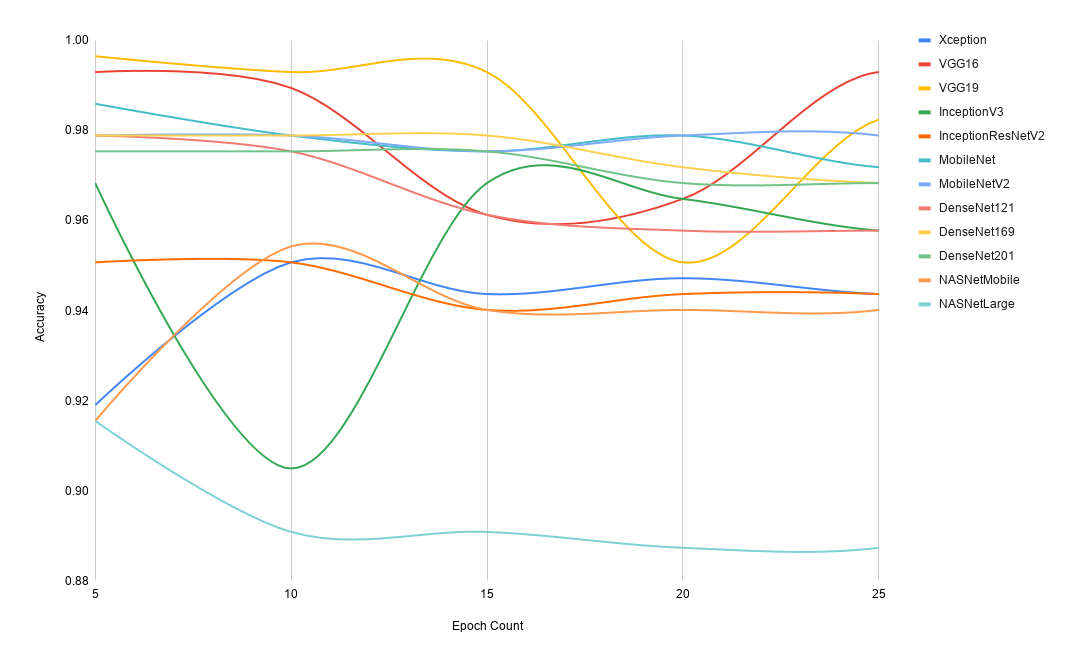}
 \caption{Accuracy of dependable models based on the 5-25 epoch range.}
 \label{fig:Results_Chart_5_to_25_Epochs_No_ResNet}
\end{figure}

These results connect well to the previous section, where we outlined the unique circumstances that surround this particular problem. We note that we are training models on extremely specific objects in space, with unique features, patterns, colors, etc. Advanced models such as ResNet and Inception learn too well and proceed to analyze the extremely minute details in the training set. Then, when asked about other images that are unique in different ways, they struggle to find the connection. Meanwhile, less advanced models, such as VGG and MobileNet variants, do not lose that generality while learning.

\subsection{Implications for future simulator training}
%Talk about how accuracy stems from just 300 or so images, easy to reproduce

Results show quite a few viable models, depending on whether we want extreme accuracy at the cost of variance (VGG variants), or dependability at the slight cost of accuracy (MobileNet variants). 

Moreover, we note that this approach yields high accuracy with relatively few training images.
With only 900 training images, we have achieved 98\% accuracy with multiple models, giving us a wide variety of options depending on the situation. Thus, for future work, even small samples of simulated images can successfully train a neural network to detect real objects in space with extremely high accuracy.

\section{Foundations for Future Work}
%%%%%%%%%%%%%%%%%%%%%%%%%%%%%%%%%%%%%%%%%%%%%%%%%%%%%%%%%%%%%%%%%%%%%%%%%%%%%

We have expanded upon \cite{Bird2020} and have shown that a small subset of simulated images can produce extremely accurate predictions of real-world planets during an interstellar journey and beyond. This paper solidifies the overall outlook on optimization methods for exoplanet detection, while introducing many ideas that will open new and exciting problems in deep space exploration. The same methodology can be used in a wide variety of astrophysical (and other) applications, where subtle issues in both the temporal and spatial domain are critical to access, and make decisions for, low bandwidth return applications. 

An upcoming paper will address categorization, which will expand the ideas of simulator-based detection to objects beyond exoplanets. In particular, we will further explore whether simulators can help train neural networks to distinguish between specific types of planets. What about specific types of stars, comets, asteroids, and even more interestingly, signs of life?

\section{Conclusion}

In our previous work, we showed how simulator images could be used to successfully train a neural network to identify real images of planets. In this paper, we delve into specific model optimization and obtain some fascinating results. First, multiple models are proven to have above 99\% accuracy when trained only on simulator images and tested on real images of planets. This result completely supports a simulator-based training model for deep space journeys, allowing us to train large neural networks pre-flight on Earth. Second, we have shown that extremely high accuracy does not depend on large data sets in this niche problem. With under 1,000 training images, we have achieved over 98\% maximum accuracy with six different models. Finally, we demonstrate that there exists both high accuracy and high stability models that can perform well with no false negatives.

\section{Acknowledgements}
%\acknowledgements
\subsection{NASA}
PML gratefully acknowledges funding from NASA NIAC
NNX15AL91G and NASA NIAC NNX16AL32G for the NASA Starlight program
and the
NASA California Space Grant NASA NNX10AT93H,
a generous gift from the Emmett and Gladys
W. Technology Fund, as well as support
from the Breakthrough Foundation for its Breakthrough StarShot
program.
% This reference needs to be updated manually
% by adding year={} field to dm.bib.
More details on the NASA Starlight program can be found
at \url{www.deepspace.ucsb.edu/Starlight}.

\subsection{NSF}

This work used the Extreme Science and Engineering Discovery Environment (XSEDE), which is supported by National Science Foundation grant number ACI-1548562. Models were run through initial testing phases using the Comet GPU cluster, allocation ID: TG-CCR180013.

%%%%%%%%%%%%%%%%%%%%%%%%%%%%%%%%%%%%%%%%%%%%%%%%%%%%%%%%%%%%%%%%%%%%%%%%%%%%%
%% Appendices
% \appendix

%\clearpage

%\begin{table}
%\caption{This is the caption of this table}
%\begin{tabular}{ll}
%\hline
%Parameter&Value\\
%\hline
%Parameter 1 & $526.849 \pm 0.003$ s \\
%Parameter 2 & $5268.49 \pm 0.03$ s \\
%Parameter 3 & $52684.9 \pm 0.3$ s \\
%\hline
%\end{tabular}
%\label{table1}
%\end{table}

\end{document}